\documentclass[submission,copyright]{eptcs}
\providecommand{\event}{ICLP/LPNMR-DC 2024} 

\usepackage{iftex}
\usepackage{graphicx}
\usepackage{listings}
\usepackage{xcolor}
\usepackage{makecell}

\lstdefinestyle{prolog}
{
    language=Prolog,
    basicstyle = \ttfamily\color{blue},
    moredelim = [s][\color{black}]{(}{)},
    literate =
        {:-}{{\textcolor{black}{:-}}}2
        {,}{{\textcolor{black}{,}}}1
        {.}{{\textcolor{black}{.}}}1,
    numbers=left,
    xleftmargin=2em,
    numberstyle=\tiny\color{gray}
}
\ifpdf
  \usepackage{underscore}         
  \usepackage[T1]{fontenc}        
\else
  \usepackage{breakurl}           
\fi

\title{Autonomous Task Completion Based on\\Goal-directed Answer Set Programming}
\author{Alexis R. Tudor
\institute{University of Texas at Dallas\\Texas, USA}
\email{alexisrenee1@gmail.com}
}
\def\titlerunning{s(CASP) for Task Completion}
\def\authorrunning{A.R. Tudor}
\begin{document}
\maketitle

\begin{abstract}
Task planning for autonomous agents has typically been done using deep learning models and simulation-based reinforcement learning. This research proposes combining inductive learning techniques with goal-directed answer set programming to increase the explainability and reliability of systems for task breakdown and completion. Preliminary research has led to the creation of a Python harness that utilizes s(CASP) to solve task problems in a computationally efficient way. Although this research is in the early stages, we are exploring solutions to complex problems in simulated task completion. 
\end{abstract}

\section{Introduction}
Task planning for autonomous agents has been an area of interest in recent years as robotics and deep learning have made major advances. Most approaches to task planning involve the use of deep learning models. The most popular approach is deep reinforcement learning, though recent work has used large-language models (LLMs) as well. Deep learning models generally achieve good results, however, they are uninterpretable and often produce flawed answers with no explanation. Much work has been done to improve the explainability of deep learning models, however, they remain untrustworthy. 

A better solution is to use logic programming. Logic programming is a programming paradigm based primarily on the calculation of Horn clauses through the process of entailment. The most common logic programming language is Prolog, though most logic programming languages consist of a Prolog-like collection of facts and rules. One advantage of logic programs is that they are inherently interpretable and their errors can be logically understood and solved. The research proposed in this paper involves using answer set programming to complete tasks in a simulated environment. This will hopefully result in autonomous task planning that is both robust and trustworthy.

\section{Background and Relevant Literature}
As autonomous agents become more ubiquitous, the focus has turned to their ability to complete complex tasks in the real world, converting high-level instructions (like "fold laundry") to executable plans ("walk to clothes", "grab clothes", etc.). Autonomous task completion can mean anything from unmanned vehicles navigating from one point to another to robotic kitchen assistants designed to make certain foods. For the most part, modern autonomous systems use deep learning models to accomplish this \cite{morales2021}. This commonly takes the form of deep reinforcement learning and more recently LLMs. Deep learning has achieved excellent results on complex problems. However, most deep learning systems are black boxes that lack explainability and interpretability. This is especially dangerous given how dependent deep learning algorithms are on the (often flawed) data they are trained on. This makes it difficult to trust that their answers are correct and unbiased, as explored in DARPA's explainable AI retrospective \cite{gunning2021}. This is important in critical systems, such as hospital diagnoses or military applications, where it has to be quickly apparent whether a model is correct or not. LLMs are vulnerable to "hallucination", where they provide incorrect responses that are statistically likely \cite{huang2023}. Additionally, they can be "jailbroken" to respond outside of the bounds they were designed for \cite{chao2023}. In the specific arena of autonomous task completion, LLMs struggle with making task breakdowns that are both correct and executable \cite{huang22}. Deep learning as a whole is well explained in other high-quality survey papers \cite{dong2021}. This paper will focus more on the importance of explainability, which is often neglected in deep learning models.

Inductive Logic Programming (ILP) is a form of machine learning that codifies its learning in the form of first-order logic. Ever since the term was defined in 1991 \cite{muggleton1991} as the "intersection
of Logic Programming and Machine Learning", ILP has served to solve machine learning problems. ILP can get results rivaling deep learning models while being inherently interpretable and explainable \cite{zhang2023}. Recent advances in ILP, such as the FOLD family of algorithms, demonstrate that complex data can be represented in small logic programs using default rules. A more detailed description of default rules and the FOLD family of incremental learning algorithms can be found in the paper by Gupta et al. \cite{gupta2023}. The research mentioned above uses a type of logic programming called Answer Set Programming (ASP). Unlike Prolog-based logic programming, which generates a true or false answer for a queried predicate, ASP is used to generate all entailable rules from a knowledge base. This collection is called an answer set. This can be used to generate "multiple worlds" where different answer sets are true.

Traditional ASP, like in Clingo  \cite{gebser2014}, executes an answer set program through the use of a SAT solver and grounding. Grounding involves the generation of the program with all variables substituted with constants in the program. A disadvantage of this approach is that grounding is not always guaranteed to be feasible, which can leave some programs with no ASP solution. The s(CASP) system  \cite{arias2018} solves this problem by performing a top-down goal-oriented search which eliminates the need for grounding. This advantage makes s(CASP) well-suited to the representation of complex world states and provides an advantage over other ASP systems \cite{gelfond2014}.

One of the biggest weaknesses of the ILP approach to solving problems is the need for background information and 'program templates'. Program templates are a layout of how the generated information should look in the context of the logic program. A domain expert must provide this program template and explicitly logic program-based background knowledge for most ILP. Thus, for trivial examples, it would be just as easy to include the final found rules in the knowledge base at the start. Additionally, while ILP programs perform very well on data that can be represented in a logic program, logic programs have a difficult time representing complex data. These weaknesses can be overcome through the use of traditional machine learning algorithms to supplement a logic program. This approach increases explainability while utilizing the benefits of deep learning and other machine learning models, such as in the paper by Rajasekharan et al. that uses an s(CASP) knowledge base to constrain an LLM into providing more reliable results \cite{rajasekharan2023}. Other examples exist of using some form of knowledge base to improve deep learning algorithms \cite{uchendu2023} \cite{hao2023}, but the use of logic programming to augment other algorithms merits further exploration.

\begin{figure*}[tp]
\centering
\includegraphics[width = 300px]{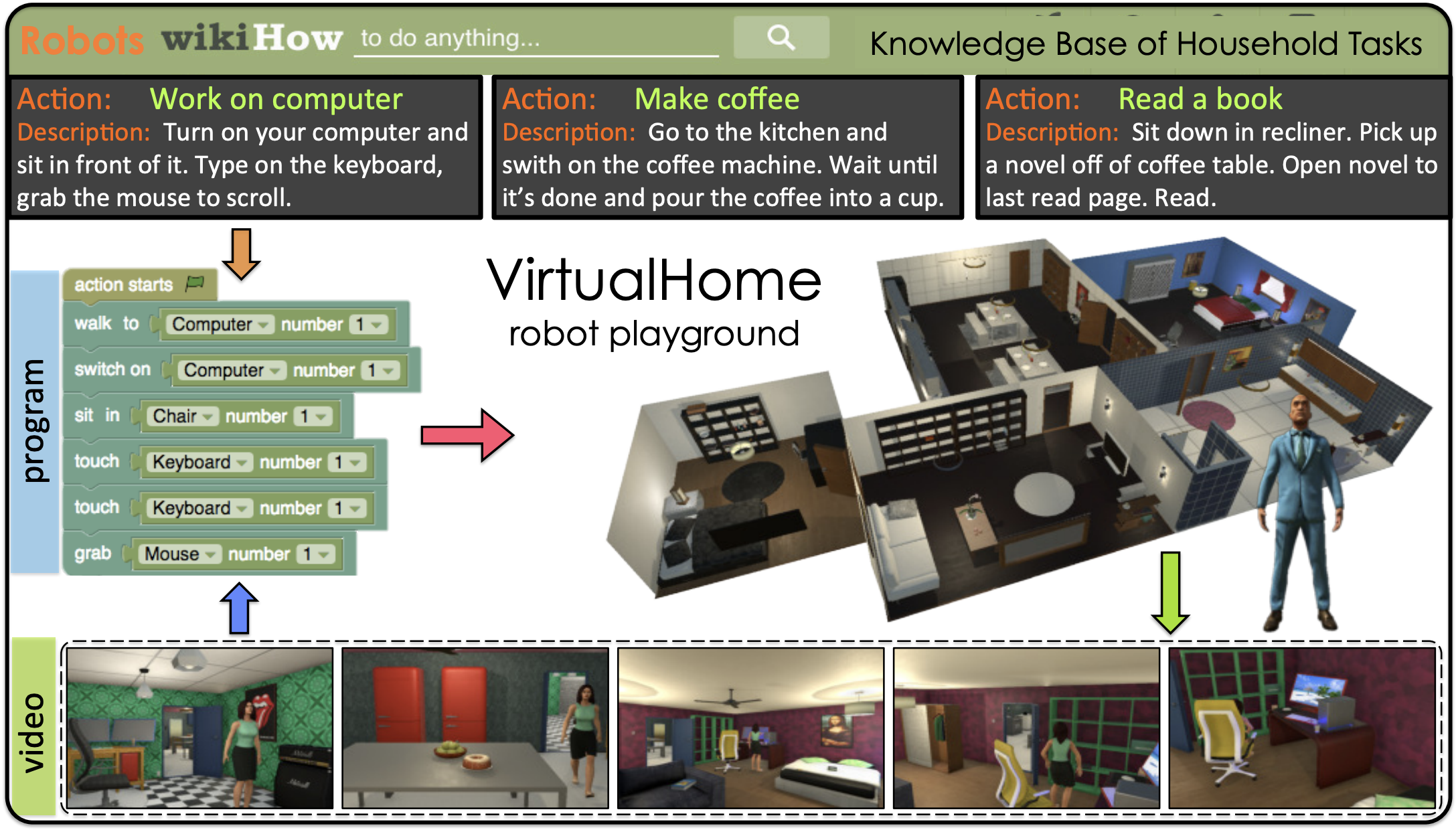}
\caption{An image showcasing the VirtualHome simulator and an example of its associated task instructions \cite{puig2018}.}
\label{fig:vh}
\end{figure*}

\section{Methodology}
The research outlined in this paper seeks to explore the use of an s(CASP) knowledge base for autonomous task completion in a simulated virtual environment. To test our system, we use the VirtualHome simulator (shown in Figure \ref{fig:vh}) as a playground for our s(CASP) agent to perform tasks in. VirtualHome allows for multiple agents to operate in a variety of simulated apartments, and provides a large database of high-level task breakdowns into step-by-step instructions. This simulation proved to be especially useful for our research because it has a "mid-level" control scheme. This means that we can give the agent commands like "grab remote" rather than dealing with the details of actual movement ("move left foot 3 inches forward", "rotate right arm 45 degrees at the elbow joint", etc) that would be more appropriate for a detailed robotic controller.  

The primary goal of this research is to achieve reasonably accurate task completion using goal-directed answer set programming. The end system would have a high level of explainability for decision-making, where the results are trustworthy and could be diagnosed if in error. We wish to further prove that even the very high-quality deep learning systems in use today could be augmented through the use of logic programming. Using logic in this way moves toward general artificial intelligence. Using s(CASP) to simulate how humans can perform common-sense logical interactions with the world brings us closer to reasoning AI. 

\begin{figure*}[htp]
\centering
\fbox{\includegraphics[width = 300px]{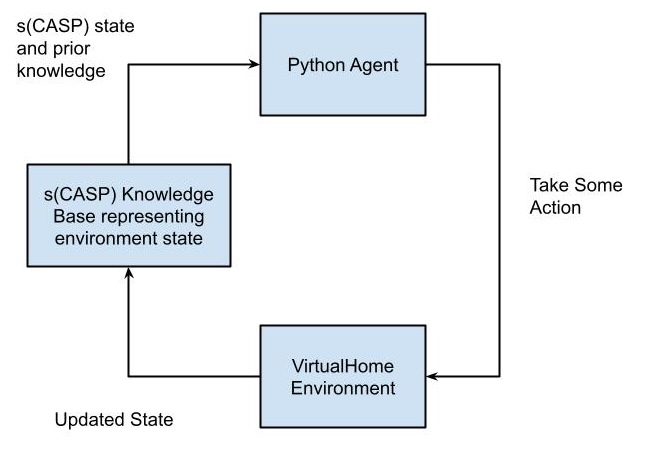}}
\caption{A diagram showing the high-level functionality of the Python harness for s(CASP). The Python harness can perform actions in the VirtualHome environment, and then convert the state of the environment to s(CASP) facts. These facts can then be used to inform the next action of the agent.}
\label{fig:pythonharness}
\end{figure*}

An additional goal of this research is to make s(CASP) easier to use with simulators. A notable weakness of s(CASP) is that it does not have a Python API, which makes it difficult to run in line with other forms of machine learning. The software engineering goal of this research is to create a "harness" for using s(CASP) in Python for interactions with simulators, as shown in Figure \ref{fig:pythonharness}.

\subsection{Status}
Although this research is still in an early stage, there have already been promising results in producing executable actions for small-scale real-world tasks. Using the Python harness mentioned above, the simulated VirtualHome environment can be instantiated and transformed into an s(CASP) representation of the world state:
\begin{lstlisting}
% With Time
current_time(1).
type(livingroom100, livingroom).
type(remotecontrol1, remotecontrol).
off(remotecontrol1, 1).
inside([inside(remotecontrol1, livingroom100), 
        inside(character0, livingroom100)], 1).
% Without time
type(livingroom100, livingroom).
type(remotecontrol1, remotecontrol).
off(remotecontrol1).
inside([inside(remotecontrol1, livingroom100),
        inside(character0, livingroom100)]).
\end{lstlisting}
The above example represents a world state containing a single turned-off remote control sitting in a living room at time 1. The Python harness keeps track of a discretized world time where each action taken by the agent represents a step forward in time, however the addition of time greatly increases the complexity of the world state s(CASP) program. Using time naively in this manner results in intractable programs which loop over infinite time, and so when representing the world state we use the latter example where timestamps are not provided in the state facts. Even without the use of time, this representation of the world state easily grows to encompass a large amount of facts. The complexity of generating an answer set that accounts for all of these facts and possible worlds quickly becomes a computational obstacle. For testing purposes, the Python harness has a small-scale simulation environment built in. Still, the goal remains to execute plans in realistic environments.

To represent and complete tasks we treat task completion as a planning problem. We represent each task as a final state (i.e. if the task was to grab a remote control, the final state would include holds(remotecontrol)) and then formulate actions to reach that final state. The added complexity to this comes from the incorporation of the simulated world state when starting from an initial state. We use the following s(CASP) rules for the task planning problem:
\\
\begin{lstlisting}
% Planning
% Get the initial state of items close to the character
initial_state(List) :- close_to_character(List).

% Find a set of actions to reach the final state
transform(FinalState, Plan) :- 
    initial_state(State1), 
    transform(State1, FinalState, [State1], Plan).
transform(State1, FinalState,_,[]) :- subset(FinalState, State1).
transform(State1, State2, Visited, [Action|Actions]) :-
    choose_action(Action, State1, State2),
    update(Action, State1, State),
    not member(State, Visited),
    transform(State, State2, [State|Visited], Actions).

% We choose an action to take
choose_action(Action, State1, State2) :- 
    suggest(Action, State2), legal_action(Action, State1).
choose_action(Action, State1, _) :- 
    legal_action(Action, State1).
suggest(walk(X), State) :- member(close(X), State).

% Check if an action is legal given the state
legal_action(walk(X), State) :- 
    type(X, Y), Y \= character, not member(close(X), State).

% Update state
update(walk(X), State, [close(X) | State1]) :- 
    update_walking(X, State, State, [], State1).

% Tasks
complete_task(walk_to_remote, P) :- 
    type(Remote, remotecontrol), transform([close(Remote)], P).
\end{lstlisting}    

These rules are a small representative subset of the rules used to generate actions to complete a task. In this very simple example, the task is to walk towards a remote control, which can be easily accomplished by the program. Using this knowledge base we can also achieve some inference. Given a final state where the agent is holding something, using the s(CASP) knowledge base constraints the agent can intuit that it first needs to walk to the item before attempting to pick it up. 

A serious problem with this inference, however, is that in sufficiently large environments it becomes too long to calculate (at least over twelve hours without concluding). For example, in the small-scale testing simulation that contains only six items, a plan for "grab the remote" can result in the agent walking to every other object in the room before walking to the remote to grab it. In addition to that solution being inefficient, it takes impossibly long in a real environment with nearly 500 objects. This problem can be solved by adding the rule \lstinline{suggest(walk(X), State) :- member(holds(X), State),} \lstinline{not member(close(X), State).}, however the same then must be done for any other state requiring closeness as a prerequisite. This decreases the value of logical inference and increases the rules required for simple task planning. The limitation remains computation time.

\subsection{Preliminary Results}
We have made significant strides in reducing the impact of computation time on the program. To reduce computation time, we implemented a dynamic dependency graph that is used to remove facts and rules that are not relevant to the query. For example, given the following knowledge base:
\begin{lstlisting}
parent(tony, abe).
parent(tony, jill).
parent(abe, sarah).

male(tony).
male(abe).
female(jill).
female(sarah).

parent(Parent, Child) :- sibling(X, Child), parent(Parent, X).
grandparent(Grandparent, Child) :- 
    parent(Grandparent, Parent), parent(Parent, Child).
sibling(X,Y) :- parent(Parent, X), parent(Parent, Y), X\=Y.
auntuncle(AU,N) :- sibling(AU, Parent), parent(Parent, N).
niece(Niece, AU) :- auntuncle(AU, Niece), female(Niece).
\end{lstlisting}

\begin{figure*}[t]
\centering
\fbox{\includegraphics[width = 200px]{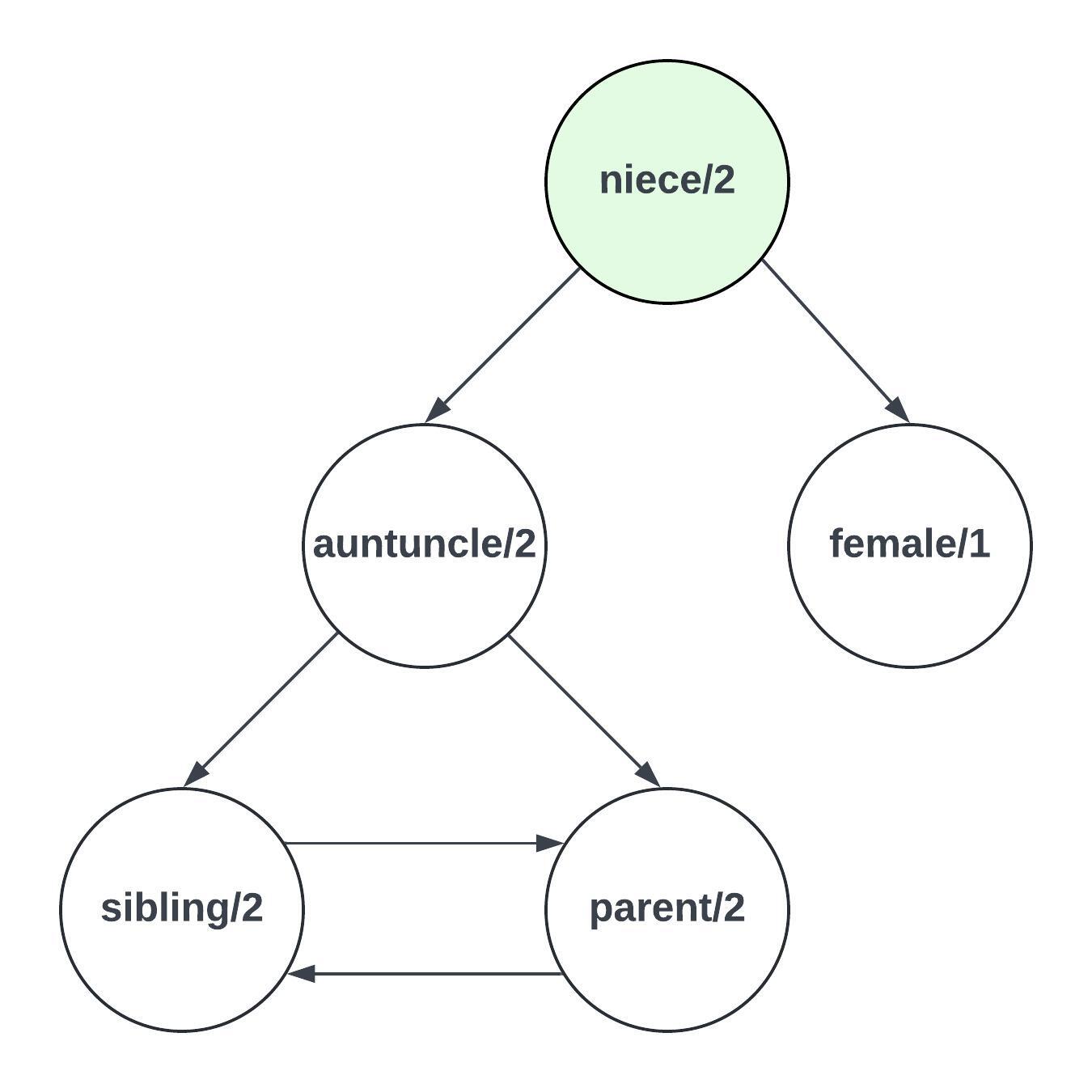}}
\caption{An example dependency graph for a family tree program where niece is the queried rule.}
\label{fig:dependencygraph}
\end{figure*}
\pagebreak
Generating a dependency graph for \lstinline{?- niece(X, Y).} produces the graph in Figure \ref{fig:dependencygraph}. Using the dependency graph the Python harness can simplify the above knowledge base, removing the male/1 and grandparent/2 predicates entirely. In a program of this size, the computational savings of such optimization is negligible. However, preliminary research has shown a significant time saving in the real-world environment. Table \ref{tab:optimization} demonstrates the time savings of using the dependency graph to prune the knowledge base for the specific task being accomplished on three semi-simple tasks that take one to four actions to fulfill. The computational time can be reduced from nearly thirty minutes to a fraction of a second using this approach, allowing for continued research into more complex tasks. 

\begin{table} [t]
\centering
\caption{Table of computational time in seconds for three tasks completed by the s(CASP) agent both optimized and unoptimized.}
\label{tab:optimization}
\begin{tabular}{ |p{3cm}||p{3cm}|p{3cm}|p{3cm}|  }
 \hline
 \makecell{Task}&\makecell{Unoptimized\\Time to Complete\\(s)} &\makecell{Dependency Graph\\Optimized Time to\\Complete (s)}\\
 \Xhline{3\arrayrulewidth}
 \makecell{Grab Remote\\Control}   & 13925.14    &0.55 \\
 \hline
 \makecell{Grab Remote\\Control and Shirt}&   608.28 & 0.71   \\
 \hline
 \makecell{Grab Cell Phone\\and Sit on Couch} &1771.21 & 0.64\\
 \hline
\end{tabular}
\end{table}

\subsection{Open Issues and Expected Achievements}
Right now, the biggest issues facing this research concern the representation of the s(CASP) knowledge base. There are several outstanding questions. 

\paragraph{Representing a Complex Real-World State} Representing a simulation of any reasonable size leads to an exponential increase in the number of facts available in the world state. In addition to these facts, there also needs to be a set of rules adequate to perform tasks in the environment. This produces answer sets that are intractable to generate. The use of a dependency graph to pare down the knowledge base allows us to perform more complicated tasks, however there can be more optimization.

Another solution that will be explored is to keep groups of state facts and rules in different programs. The creation of modules that correspond to various tasks or locations would allow for faster calculation of relevant queries. This follows the human logic that one likely does not need their cooking knowledge if, for example, they need to walk their dog. 

\paragraph{The Passage of Time} As mentioned above, the use of time in the knowledge base provides complications related to the ostensibly infinitely divisible nature of time (as posited by the famous Greek philosopher Zeno). This is a known problem with representing continuous time in logic programming and would require the inclusion of event calculus \cite{varanasi2022}. 

\paragraph{Large-scale Learning} As deep learning and its applications for real-world task completion are already well explored, the value of this research lies in seeing how complex problems that the s(CASP) task planner can solve can get. To that end, explanation-based learning is a promising paradigm that would allow for generalized knowledge from a small number of examples \cite{strout2019} and works well with answer set programming. 

Likely, s(CASP) by itself cannot encode all of the complexities of a real environment and remain tractable. Once that point is reached, there would still be benefits in combining s(CASP) with more traditional machine learning (and newer deep learning, such as LLMs) to improve performance in the former and explainability of the latter. We hope to leverage databases of task instructions and breakdowns, such as those provided by VirtualHome or ALFRED \cite{Shridhar2019}, to improve the performance of the s(CASP) agent at scale. 
\\
\\
We expect to be able to answer these questions in a unified way to facilitate task completion in complex environments using s(CASP). Although solutions to these problems may always become intractable at certain levels of fidelity, there is valuable knowledge to be gained along the way.

\section{Conclusion}
In conclusion, this line of research could open up a broad number of solutions for challenging ILP problems. Simply creating a Python framework for the use of s(CASP) with simulated environments is an advancement for s(CASP), as it is currently lacking a Python API. Using the intersection of ILP and traditional machine learning is promising for improving the explainability and reliability of task-completing autonomous agents. 

\nocite{*}
\bibliographystyle{eptcs}
\bibliography{generic}

\begin{thebibliography}{10}
\providecommand{\bibitemdeclare}[2]{}
\providecommand{\surnamestart}{}
\providecommand{\surnameend}{}
\providecommand{\urlprefix}{Available at }
\providecommand{\url}[1]{\texttt{#1}}
\providecommand{\href}[2]{\texttt{#2}}
\providecommand{\urlalt}[2]{\href{#1}{#2}}
\providecommand{\doi}[1]{doi:\urlalt{https://doi.org/#1}{#1}}
\providecommand{\eprint}[1]{arXiv:\urlalt{https://arxiv.org/abs/#1}{#1}}
\providecommand{\bibinfo}[2]{#2}

\bibitemdeclare{article}{arias2018}
\bibitem{arias2018}
\bibinfo{author}{Joaqin \surnamestart Arias\surnameend},
  \bibinfo{author}{Manuel \surnamestart Carro\surnameend},
  \bibinfo{author}{Elmer \surnamestart Salazar\surnameend},
  \bibinfo{author}{Kyle \surnamestart Marple\surnameend} \&
  \bibinfo{author}{Gopal \surnamestart Gupta\surnameend}
  (\bibinfo{year}{2018}): \emph{\bibinfo{title}{Constraint Answer Set
  Programming without Grounding}}.
\newblock {\slshape \bibinfo{journal}{Theory and Practice of Logic
  Programming}} \bibinfo{volume}{18}(\bibinfo{number}{3-4}), p.
  \bibinfo{pages}{337–354}, \doi{10.1017/S1471068418000285}.

\bibitemdeclare{misc}{chao2023}
\bibitem{chao2023}
\bibinfo{author}{Patrick \surnamestart Chao\surnameend},
  \bibinfo{author}{Alexander \surnamestart Robey\surnameend},
  \bibinfo{author}{Edgar \surnamestart Dobriban\surnameend},
  \bibinfo{author}{Hamed \surnamestart Hassani\surnameend},
  \bibinfo{author}{George~J. \surnamestart Pappas\surnameend} \&
  \bibinfo{author}{Eric \surnamestart Wong\surnameend} (\bibinfo{year}{2023}):
  \emph{\bibinfo{title}{Jailbreaking Black Box Large Language Models in Twenty
  Queries}}.
\newblock \eprint{2310.08419}.

\bibitemdeclare{article}{dong2021}
\bibitem{dong2021}
\bibinfo{author}{Shi \surnamestart Dong\surnameend}, \bibinfo{author}{Ping
  \surnamestart Wang\surnameend} \& \bibinfo{author}{Khushnood \surnamestart
  Abbas\surnameend} (\bibinfo{year}{2021}): \emph{\bibinfo{title}{A survey on
  deep learning and its applications}}.
\newblock {\slshape \bibinfo{journal}{Computer Science Review}}
  \bibinfo{volume}{40}, p. \bibinfo{pages}{100379},
  \doi{10.1016/j.cosrev.2021.100379}.
\newblock
  \urlprefix\url{https://www.sciencedirect.com/science/article/pii/S1574013721000198}.

\bibitemdeclare{article}{gebser2014}
\bibitem{gebser2014}
\bibinfo{author}{M.~\surnamestart Gebser\surnameend}, \bibinfo{author}{Roland
  \surnamestart Kaminski\surnameend}, \bibinfo{author}{Benjamin \surnamestart
  Kaufmann\surnameend} \& \bibinfo{author}{Torsten \surnamestart
  Schaub\surnameend} (\bibinfo{year}{2014}): \emph{\bibinfo{title}{Clingo = ASP
  + Control: Preliminary Report}}.
\newblock {\slshape \bibinfo{journal}{ArXiv}} \bibinfo{volume}{abs/1405.3694},
  \doi{10.48550/arXiv.1405.3694}.
\newblock \urlprefix\url{http://arxiv.org/abs/1405.3694}.

\bibitemdeclare{book}{gelfond2014}
\bibitem{gelfond2014}
\bibinfo{author}{Michael \surnamestart Gelfond\surnameend} \&
  \bibinfo{author}{Yulia \surnamestart Kahl\surnameend} (\bibinfo{year}{2014}):
  \emph{\bibinfo{title}{Knowledge representation, reasoning, and the design of
  Intelligent Agents: The answer-set programming approach}}.
\newblock \bibinfo{publisher}{Cambridge University Press},
  \doi{10.1017/CBO9781139342124}.

\bibitemdeclare{article}{gunning2021}
\bibitem{gunning2021}
\bibinfo{author}{David \surnamestart Gunning\surnameend}, \bibinfo{author}{Eric
  \surnamestart Vorm\surnameend}, \bibinfo{author}{Jennifer~Yunyan
  \surnamestart Wang\surnameend} \& \bibinfo{author}{Matt \surnamestart
  Turek\surnameend} (\bibinfo{year}{2021}): \emph{\bibinfo{title}{DARPA's
  explainable AI (XAI) program: A retrospective}}.
\newblock {\slshape \bibinfo{journal}{Applied AI Letters}}
  \bibinfo{volume}{2}(\bibinfo{number}{4}), p. \bibinfo{pages}{e61},
  \doi{10.1002/ail2.61}.
\newblock \eprint{https://onlinelibrary.wiley.com/doi/pdf/10.1002/ail2.61}.

\bibitemdeclare{article}{gupta2023}
\bibitem{gupta2023}
\bibinfo{author}{Gopal \surnamestart Gupta\surnameend}, \bibinfo{author}{Huaduo
  \surnamestart Wang\surnameend}, \bibinfo{author}{Kinjal \surnamestart
  Basu\surnameend}, \bibinfo{author}{Farhad \surnamestart Shakerin\surnameend},
  \bibinfo{author}{Elmer \surnamestart Salazar\surnameend},
  \bibinfo{author}{Sarat~Chandra \surnamestart Varanasi\surnameend},
  \bibinfo{author}{Parth \surnamestart Padalkar\surnameend} \&
  \bibinfo{author}{Sopam \surnamestart Dasgupta\surnameend}
  (\bibinfo{year}{2023}): \emph{\bibinfo{title}{Logic-based explainable and
  incremental machine learning}}.
\newblock {\slshape \bibinfo{journal}{Prolog: The Next 50 Years}}, p.
  \bibinfo{pages}{346–358}, \doi{10.1007/978-3-031-35254-6_28}.

\bibitemdeclare{misc}{hao2023}
\bibitem{hao2023}
\bibinfo{author}{Zhongkai \surnamestart Hao\surnameend},
  \bibinfo{author}{Songming \surnamestart Liu\surnameend},
  \bibinfo{author}{Yichi \surnamestart Zhang\surnameend},
  \bibinfo{author}{Chengyang \surnamestart Ying\surnameend},
  \bibinfo{author}{Yao \surnamestart Feng\surnameend}, \bibinfo{author}{Hang
  \surnamestart Su\surnameend} \& \bibinfo{author}{Jun \surnamestart
  Zhu\surnameend} (\bibinfo{year}{2023}):
  \emph{\bibinfo{title}{Physics-Informed Machine Learning: A Survey on
  Problems, Methods and Applications}}.
\newblock \eprint{2211.08064}.

\bibitemdeclare{misc}{huang2023}
\bibitem{huang2023}
\bibinfo{author}{Lei \surnamestart Huang\surnameend}, \bibinfo{author}{Weijiang
  \surnamestart Yu\surnameend}, \bibinfo{author}{Weitao \surnamestart
  Ma\surnameend}, \bibinfo{author}{Weihong \surnamestart Zhong\surnameend},
  \bibinfo{author}{Zhangyin \surnamestart Feng\surnameend},
  \bibinfo{author}{Haotian \surnamestart Wang\surnameend},
  \bibinfo{author}{Qianglong \surnamestart Chen\surnameend},
  \bibinfo{author}{Weihua \surnamestart Peng\surnameend},
  \bibinfo{author}{Xiaocheng \surnamestart Feng\surnameend},
  \bibinfo{author}{Bing \surnamestart Qin\surnameend} \& \bibinfo{author}{Ting
  \surnamestart Liu\surnameend} (\bibinfo{year}{2023}): \emph{\bibinfo{title}{A
  Survey on Hallucination in Large Language Models: Principles, Taxonomy,
  Challenges, and Open Questions}}.
\newblock \eprint{2311.05232}.

\bibitemdeclare{inproceedings}{huang22}
\bibitem{huang22}
\bibinfo{author}{Wenlong \surnamestart Huang\surnameend},
  \bibinfo{author}{Pieter \surnamestart Abbeel\surnameend},
  \bibinfo{author}{Deepak \surnamestart Pathak\surnameend} \&
  \bibinfo{author}{Igor \surnamestart Mordatch\surnameend}
  (\bibinfo{year}{2022}): \emph{\bibinfo{title}{Language Models as Zero-Shot
  Planners: Extracting Actionable Knowledge for Embodied Agents}}.
\newblock In \bibinfo{editor}{Kamalika \surnamestart Chaudhuri\surnameend},
  \bibinfo{editor}{Stefanie \surnamestart Jegelka\surnameend},
  \bibinfo{editor}{Le~\surnamestart Song\surnameend}, \bibinfo{editor}{Csaba
  \surnamestart Szepesvari\surnameend}, \bibinfo{editor}{Gang \surnamestart
  Niu\surnameend} \& \bibinfo{editor}{Sivan \surnamestart Sabato\surnameend},
  editors: {\slshape \bibinfo{booktitle}{Proceedings of the 39th International
  Conference on Machine Learning}}, {\slshape \bibinfo{series}{Proceedings of
  Machine Learning Research}} \bibinfo{volume}{162}, \bibinfo{publisher}{PMLR},
  pp. \bibinfo{pages}{9118--9147}, \doi{10.48550/arXiv.2201.07207}.
\newblock \urlprefix\url{https://proceedings.mlr.press/v162/huang22a.html}.

\bibitemdeclare{article}{morales2021}
\bibitem{morales2021}
\bibinfo{author}{Eduardo~F \surnamestart Morales\surnameend},
  \bibinfo{author}{Rafael \surnamestart Murrieta-Cid\surnameend},
  \bibinfo{author}{Israel \surnamestart Becerra\surnameend} \&
  \bibinfo{author}{Marco~A \surnamestart Esquivel-Basaldua\surnameend}
  (\bibinfo{year}{2021}): \emph{\bibinfo{title}{A survey on deep learning and
  deep reinforcement learning in robotics with a tutorial on deep reinforcement
  learning}}.
\newblock {\slshape \bibinfo{journal}{Intelligent Service Robotics}}
  \bibinfo{volume}{14}(\bibinfo{number}{5}), pp. \bibinfo{pages}{773--805},
  \doi{10.1007/s11370-021-00398-z}.

\bibitemdeclare{article}{muggleton1991}
\bibitem{muggleton1991}
\bibinfo{author}{Stephen \surnamestart Muggleton\surnameend}
  (\bibinfo{year}{1991}): \emph{\bibinfo{title}{Inductive logic programming}}.
\newblock {\slshape \bibinfo{journal}{New Generation Computing}}
  \bibinfo{volume}{8}(\bibinfo{number}{4}), p. \bibinfo{pages}{295–318},
  \doi{10.1007/bf03037089}.

\bibitemdeclare{misc}{puig2018}
\bibitem{puig2018}
\bibinfo{author}{Xavier \surnamestart Puig\surnameend}, \bibinfo{author}{Kevin
  \surnamestart Ra\surnameend}, \bibinfo{author}{Marko \surnamestart
  Boben\surnameend}, \bibinfo{author}{Jiaman \surnamestart Li\surnameend},
  \bibinfo{author}{Tingwu \surnamestart Wang\surnameend},
  \bibinfo{author}{Sanja \surnamestart Fidler\surnameend} \&
  \bibinfo{author}{Antonio \surnamestart Torralba\surnameend}
  (\bibinfo{year}{2018}): \emph{\bibinfo{title}{VirtualHome: Simulating
  Household Activities via Programs}}.
\newblock \eprint{1806.07011}.

\bibitemdeclare{article}{rajasekharan2023}
\bibitem{rajasekharan2023}
\bibinfo{author}{Abhiramon \surnamestart Rajasekharan\surnameend},
  \bibinfo{author}{Yankai \surnamestart Zeng\surnameend},
  \bibinfo{author}{Parth \surnamestart Padalkar\surnameend} \&
  \bibinfo{author}{Gopal \surnamestart Gupta\surnameend}
  (\bibinfo{year}{2023}): \emph{\bibinfo{title}{Reliable Natural Language
  Understanding with Large Language Models and Answer Set Programming}}.
\newblock {\slshape \bibinfo{journal}{Electronic Proceedings in Theoretical
  Computer Science}} \bibinfo{volume}{385}, pp. \bibinfo{pages}{274--287},
  \doi{10.4204/EPTCS.385.27}.

\bibitemdeclare{article}{Shridhar2019}
\bibitem{Shridhar2019}
\bibinfo{author}{Mohit \surnamestart Shridhar\surnameend},
  \bibinfo{author}{Jesse \surnamestart Thomason\surnameend},
  \bibinfo{author}{Daniel \surnamestart Gordon\surnameend},
  \bibinfo{author}{Yonatan \surnamestart Bisk\surnameend},
  \bibinfo{author}{Winson \surnamestart Han\surnameend},
  \bibinfo{author}{Roozbeh \surnamestart Mottaghi\surnameend},
  \bibinfo{author}{Luke \surnamestart Zettlemoyer\surnameend} \&
  \bibinfo{author}{Dieter \surnamestart Fox\surnameend} (\bibinfo{year}{2019}):
  \emph{\bibinfo{title}{ALFRED: A Benchmark for Interpreting Grounded
  Instructions for Everyday Tasks}}.
\newblock {\slshape \bibinfo{journal}{2020 IEEE/CVF Conference on Computer
  Vision and Pattern Recognition (CVPR)}}, pp. \bibinfo{pages}{10737--10746},
  \doi{10.1109/cvpr42600.2020.01075}.
\newblock \urlprefix\url{https://api.semanticscholar.org/CorpusID:208617407}.

\bibitemdeclare{inproceedings}{strout2019}
\bibitem{strout2019}
\bibinfo{author}{Julia \surnamestart Strout\surnameend},
  \bibinfo{author}{Ye~\surnamestart Zhang\surnameend} \&
  \bibinfo{author}{Raymond \surnamestart Mooney\surnameend}
  (\bibinfo{year}{2019}): \emph{\bibinfo{title}{Do Human Rationales Improve
  Machine Explanations?}}
\newblock In: {\slshape \bibinfo{booktitle}{Proceedings of the 2019 ACL
  Workshop BlackboxNLP: Analyzing and Interpreting Neural Networks for NLP}},
  \bibinfo{publisher}{Association for Computational Linguistics},
  \bibinfo{address}{Florence, Italy}, pp. \bibinfo{pages}{56--62},
  \doi{10.18653/v1/W19-4807}.
\newblock \urlprefix\url{https://aclanthology.org/W19-4807}.

\bibitemdeclare{misc}{uchendu2023}
\bibitem{uchendu2023}
\bibinfo{author}{Ikechukwu \surnamestart Uchendu\surnameend},
  \bibinfo{author}{Ted \surnamestart Xiao\surnameend}, \bibinfo{author}{Yao
  \surnamestart Lu\surnameend}, \bibinfo{author}{Banghua \surnamestart
  Zhu\surnameend}, \bibinfo{author}{Mengyuan \surnamestart Yan\surnameend},
  \bibinfo{author}{Joséphine \surnamestart Simon\surnameend},
  \bibinfo{author}{Matthew \surnamestart Bennice\surnameend},
  \bibinfo{author}{Chuyuan \surnamestart Fu\surnameend}, \bibinfo{author}{Cong
  \surnamestart Ma\surnameend}, \bibinfo{author}{Jiantao \surnamestart
  Jiao\surnameend}, \bibinfo{author}{Sergey \surnamestart Levine\surnameend} \&
  \bibinfo{author}{Karol \surnamestart Hausman\surnameend}
  (\bibinfo{year}{2023}): \emph{\bibinfo{title}{Jump-Start Reinforcement
  Learning}}.
\newblock \eprint{2204.02372}.

\bibitemdeclare{inproceedings}{varanasi2022}
\bibitem{varanasi2022}
\bibinfo{author}{Sarat~Chandra \surnamestart Varanasi\surnameend},
  \bibinfo{author}{Joaqu\'{\i}n \surnamestart Arias\surnameend},
  \bibinfo{author}{Elmer \surnamestart Salazar\surnameend},
  \bibinfo{author}{Fang \surnamestart Li\surnameend}, \bibinfo{author}{Kinjal
  \surnamestart Basu\surnameend} \& \bibinfo{author}{Gopal \surnamestart
  Gupta\surnameend} (\bibinfo{year}{2022}): \emph{\bibinfo{title}{Modeling and
  Verification of Real-Time Systems with the Event Calculus and s(CASP)}}.
\newblock In: {\slshape \bibinfo{booktitle}{Practical Aspects of Declarative
  Languages: 24th International Symposium, PADL 2022, Philadelphia, PA, USA,
  January 17–18, 2022, Proceedings}}, \bibinfo{publisher}{Springer-Verlag},
  \bibinfo{address}{Berlin, Heidelberg}, p. \bibinfo{pages}{181–190},
  \doi{10.1007/978-3-030-94479-7_12}.

\bibitemdeclare{misc}{zhang2023}
\bibitem{zhang2023}
\bibinfo{author}{Zheng \surnamestart Zhang\surnameend},
  \bibinfo{author}{Liangliang \surnamestart Xu\surnameend},
  \bibinfo{author}{Levent \surnamestart Yilmaz\surnameend} \&
  \bibinfo{author}{Bo~\surnamestart Liu\surnameend} (\bibinfo{year}{2023}):
  \emph{\bibinfo{title}{A Critical Review of Inductive Logic Programming
  Techniques for Explainable AI}}.
\newblock \eprint{2112.15319}.

\end{thebibliography}
\end{document}